\documentclass{article} 
\usepackage{iclr2015,times}
\usepackage{hyperref}
\usepackage{url}
\usepackage{graphicx}
\usepackage{caption}
\usepackage{subcaption}
\usepackage{amsmath}
\usepackage{amsfonts}
\usepackage{pbox}
\usepackage{lscape}
\usepackage{multirow}
\usepackage{algorithm}
\usepackage{algpseudocode}

\title{Representation Learning for cold-start \\recommendation}

\author{
Gabriella Contardo$^*$, Ludovic Denoyer$^*$,Thierry Arti\`eres$^\phi$ \\
$^*$ Sorbonne Universit\'es, UPMC Univ Paris 06, UMR 7606, LIP6, F-75005, Paris, France \\
$^\phi$ Ecole Centrale Marseille - Laboratoire d'Informatique Fondamentale (Aix-Marseille Univ.), France \\
}

\iclrfinalcopy 

\begin{document}

\maketitle

\vspace{-0.5cm}
 \begin{abstract}


A standard approach to Collaborative Filtering (CF), i.e. prediction of user ratings on items, relies on Matrix Factorization techniques. Representations for both users and items are computed from the observed ratings and used for prediction. Unfortunatly, these transductive approaches cannot handle the case of new users arriving in the system, with no known rating, a problem known as user cold-start. A common approach in this context is to ask these incoming users for a few initialization ratings. This paper presents a model to tackle this twofold problem of (i) finding good questions to ask, (ii) building efficient representations from this small amount of information. The model can also be used in a more standard (\textit{warm}) context. Our approach is evaluated on the classical CF problem and on the cold-start problem on four different datasets showing its ability to improve baseline performance in both cases. 
%
%

\end{abstract}

  \section{Introduction}
 Most of the successful machine learning algorithms rely on data representation, i.e a way to disentangle and extract useful information from the data, which will help the model in its objective task. As highlighted by \cite{bengioRepOverview}, designing models able to learn these representations from (raw) data instead of manual pre-processing seems crucial to go further in Artificial Intelligence, and \textit{representation learning} has gain a surge of interest in machine learning. 
In parallel, recommender systems have became an active field of research and are now used in an increasing variety of applications, such as e-commerce, social networks or participative platforms. They aim to suggest the most relevant items (e.g products) to each user, in order to facilitate their experience. To recommend such relevant items, recommender systems can rely on different types of data, such as users' explicit and/or implicit feedbacks (e.g rating a movie on a scale of stars, buying an item or listening to a song), or informative features about users (age, post code) or items (type of movie, actors). One of the most common approach to recommendation is \textbf{Collaborative Filtering} (CF) which consists in making recommendation only based on the ratings provided by users over a set of items (i.e without using any additional features). 

Within CF context, a popular and efficient family of methods are \textit{Latent Factor Models}, which rely on matrix factorization-based techniques \footnote{Other families of approaches are detailed in Section \ref{rw}.}. These approaches treat the recommender problem as a representation learning one, by computing representations for users and items in a common latent space. More formally, let us consider a set $\mathcal{U}$ of $U$ known users and a set $\mathcal{I}$ of $I$ items. Let $r_{u,i}$ denote the rating of user $u \in \mathcal{U}$ for item $i \in \mathcal{I}$. A rating is usually a discrete value between 1 and 5, that can be binarized (-1/1) with a proper threshold (often 3). The $U \times I$ matrix $\mathcal{R}=\{r_{u,i}\}$ is the rating matrix which is incomplete since all ratings are not known. We will denote $\mathcal{O}$ the set of observed pairs $(u,i)$ such that a rating on item $i$ has been made by user $u$. Let us denote $N$ the dimension of the latent representation space of users and items, $p_u \in \mathbb{R}^N$ being the (learned) representation of user $u$ and $q_i \in \mathbb{R}^N$ denoting the (learned) representation of item $i$. Given these representations, classical approaches are able to compute missing ratings made by a user $u$ over an item $i$ as the dot product between $p_u$ and $q_i$. In other words, the more similar the user and the item representations are, the higher the predicted rating will be. Let us denote $\tilde{r}_{u,i}$ this 
predicted rating, we have:
\begin{equation}
\tilde{r}_{u,i}= q_i^T p_u
\end{equation}

The representation $p_u$ and $q_i$ are usually learned on the sparse input rating matrix $\mathcal{R}$ by minimizing an objective loss function over $\mathcal{L}(\mathbf{p},\mathbf{q})$ which measures the difference between observed ratings $r_{u,i}$ and predicted ones. $\mathbf{p}$ is the set of users representations, $\mathbf{q}$ being the representation of items. The loss is usually defined as a $L_2$ objective:
\begin{equation}
\mathcal{L}(\mathbf{p},\mathbf{q}) = \sum_{(u,i)\in \mathcal{O}} (r_{u,i} - q_i^T p_u)^2 + \lambda(\sum_i ||q_i||^2 + \sum_u ||p_u||^2)
\label{cfeqloss}
\end{equation}


The coefficient $\lambda$ is a manually defined regularization coefficient. This loss corresponds to a matrix decomposition in latent factors and different optimization algorithms have been proposed as alternated least squares or stochastic gradient descent (\cite{koren2009matrix}). Note that this models is a \textbf{transductive model} since it allows one to compute representations over a set of \textit{a priori} known users and items.  


The transductive nature of Matrix Factorization approaches makes them well adapted when the sets of users and of items are fixed. Yet in practical applications, new items and new users regularly appear in the system. This requires often retraining the whole system which is time consuming and also makes the system behavior unstable. Furthermore, one main limitation of transductive Matrix Factorization approaches is that they strongly rely on a certain amount of data to build relevant representations, e.g. one must have enough ratings from a new user to construct an accurate representation. Indeed, facing new users, MF methods (and more generally CF-based approaches) have to wait for this user to interact with the system and to provide ratings before being able to make recommendations for this user. These methods are thus not well-suited to propose recommendation at the beginning of the process.   

We propose to focus on the user cold-start problem\footnote{The integration of new items which is less critical in practical applications is not the focus of this paper but is discussed in the conclusion.} by interview method which consists in building a set of items on which ratings are asked to any new user. Then, recommendations are made based on this list of (incomplete) ratings. We consider a representation-learning approach which is an original approach in this context and which simultaneously learns which items to use in the interview, but also how to use these ratings for building relevant user representations. Our method is based on an inductive  model whose principle is to code ratings on items as translations in the latent representation space, allowing to easily integrate different opinions at a low computational cost. The contributions of this paper are thus the following: (i) We propose a generic representation-learning formalism for user cold-start recommendation. This formalism integrates the representation building function as part of the objective loss, and restriction over the number of items to consider in the interview process. (ii) We present a particular representation-learning model called \textbf{Inductive Additive Model} (IAM) which is based on simple assumptions about the nature of users' representations to build and that we are able to optimize using classical gradient-descent algorithms. (iii) We perform experiments on four datasets in the classical CF context as well as in the user cold-start context. Quantitative results show the effectiveness of our approach in both contexts while qualitative results show the relevancy of learned representations.

The paper is organized as follow: in Section \ref{model}, we propose the generic formulation of the representation learning problem for user cold-start, and the particular instance of model we propose. The Section \ref{exp} presents the experiments and Section \ref{rw} discusses the related work in the collaborative filtering domain. Section \ref{conclusion} proposes perspectives to this contribution.


\section{Proposed Approach}
\label{model}

We now rewrite the objective function detailed in Equation \ref{cfeqloss} in a more general form that will allow us to integrate the user cold-start problem as a representation-learning problem. As seen above, we still consider that each item will have its own learned representation denoted $q_i \in \mathbb{R}^N$ and focus on building a user representation.
When facing any new user, our model will first collect a set of ratings by asking a set of queries during an interview process. This process is composed by a set of items\footnote{The article focuses on a static interview process – i.e interview where the set of items is the same for all incoming users. A discussion on that point is provided in Section \ref{rw}} that are selected during the training phase. For each item in the interview, the new user can provide a rating, but can also choose not to provide this rating when he has no opinion. This is typically the case for example with recommendation of movies, where 
users are only able to provide ratings on movies they have seen. The model will thus have to both select relevant items to include in the interview, but also to learn how (incomplete) collected ratings will be used to build a user representation. 
Let us denote $\mathcal{Q} \subset \mathcal{I}$ the subset of items that will be used in the interview. The representation of a new incoming user $u$ will thus depend on the ratings of $u$ over $\mathcal{Q}$ that we will note $\mathcal{Q}(u)$. This representation will be given by a function $f_\Psi(\mathcal{Q}(u))$ whose parameters, to be optimized, are denoted $\Psi$. These $\Psi$ parameters are global, i.e shared by all users. The objective function of the cold-start problem (finding the parameters $\Psi$, the items' representations and the interview questions conjointly) can then be written as:
\begin{equation}
\mathcal{L}^{cold}(\mathbf{q},\Psi,\mathcal{Q}) = \sum_{(u,i)\in \mathcal{O}} (r_{u,i} - q_i^T f_\Psi(\mathcal{Q}(u)) )^2 + \lambda_1(\sum_i ||q_i||^2 + \sum_u ||f_\Psi(\mathcal{Q}(u))||^2) + \lambda_2 \#\mathcal{Q} 
\label{eql}
\end{equation}

The difference between this loss and the classical CF loss is twofold: (i) first, the learned representations $p_u$ are not free parameters, but computed by using a parametric function $f_\Psi(\mathcal{Q}(u))$, whose parameters $\Psi$ are learned; (ii) the loss includes an additional term $\lambda_2 \#\mathcal{Q} $ which measures the balance between the quality of the prediction, and the size of the interview, $\#\mathcal{Q}$ denoting the number of items of the interview; $\lambda_1$ and $\lambda_2$ are manually chosen hyper-parameters - by changing their values, the user can obtain more robust models, and models with more or less interview questions. Note that solving this problem aims at simultaneously learning the items representations, the set of items in the interview, and the parameters of the representation building function. 

%

\subsection{Inductive Additive Model (IAM)}
 
The generic formulation presented above cannot easily be optimized with any representation function. Particularly, the use of a transductive model in this context is not trivial and, when using MF-based approaches in that case, we only obtained very complex solutions with a high computation complexity. We thus need to use a more appropriate representation-learning function $f_\Psi$ that is described below.
%
%
The Inductive Additive Model (IAM) is based on two simple ideas concerning the representation of users we want to build: (i) First, one has to be able to provide good recommendation to any user that does not provide ratings during the interview process $\mathcal{Q}$.  (ii) Second we want  the user representation to be easily enriched as new ratings are available. This feature makes our approach suitable for the particular cold-start setting but also for the standard CF setting as well. 

Based on the first idea, IAM considers that any user without answers will be mapped to a representation denoted $\Psi_0 \in \mathbb{R}^N$. Moreover, the second idea naturally led us to build an additive model where a user representation is defined as a sum of the particular items representations. This means that providing a rating will yield a \textbf{translation} of the user representation in the latent space. This translation will depend on the item $i$ but also on the rating value. This translation will be learned for each possible rating value and item and denoted $\Psi^r_i$ where $r$ is the value of the rating. More precisely, in case of binary ratings \textit{like} and \textit{dislike}, the \textit{like} over a particular item will correspond to a particular translation $\Psi_i^{+1}$, and a \textit{dislike} to the translation $\Psi_i^{-1}$. The fact that the two rating values correspond to two different unrelated translations is interesting since, for some items, the \textit{dislike} rating can provide no additional information represented by a null translation, while the \textit{like} rating can be very informative, modifying the user representation - see Section \ref{exp} for a qualitative study over $\Psi$. The resulting  model $f_\Psi$ can thus be written as:
\begin{equation}
f_\Psi(u,\mathcal{Q})=\Psi_0 + \sum\limits_{(u,i) \in \mathcal{O}/ i \in \mathcal{Q}} \Psi_i^{r_{u,i}}
\label{eqiam}
\end{equation}
where the set $\{(u,i) \in \mathcal{O}/ i \in \mathcal{Q}\}$ is the set of items selected in the interview  on which user $u$ has provided a rating.

\subsubsection{Continuous Learning Problem}

Now, let us describe how the objective function described in Equation \ref{eql} with IAM model described in Equation \ref{eqiam} can be optimized. The optimization problem consisting in minimizing $\mathcal{L}^{cold}(\mathbf{q},\Psi,\mathcal{Q})$ over $\mathbf{q},\Psi$ and $\mathcal{Q}$ is a combinatorial problem since $\mathcal{Q}$ is a subset of the items. This combinatorial nature prevents us from using classical optimization methods such as gradient-descent methods and involves an intractable number of possible combinations of items. We propose to use a $L_1$ relaxation in order to transform this problem in a continuous one. Let us denote $\mathbf{\alpha} \in \mathbb{R}^I$ a weight vector, one weight per item, such that if $\alpha_i=0$ then item $i$ will not be in the interview. The cold-start loss can be rewritten with $\mathbf{\alpha}$'s as:
\begin{equation}
\mathcal{L}^{cold}(\mathbf{q},\Psi,\mathbf{\alpha}) = \sum_{(u,i)\in \mathcal{O}} (r_{u,i} - q_i^T f_\Psi(u,\mathbf{\alpha}) )^2 + \lambda |\mathbf{\alpha}|
\end{equation}
Note that the $L_2$ regularization term over the computed representation of users and items is removed here for sake of clarity. The representation of a user thus depends on the ratings made by this user for items $i$ that have a non-null weight $\alpha_i$, restricting our model to compute its prediction on a subset of items which compose the interview. If we rewrite the proposed model as:
\begin{equation}
f_\Psi(u,\mathbf{\alpha})=\Psi_0 + \sum\limits_{(u,i) \in \mathcal{O}} \alpha_i \Psi_i^{r_{u,i}}
\end{equation} 
then we obtain the following loss function:
\begin{equation}
\mathcal{L}^{cold}(\mathbf{q},\Psi,\mathbf{\alpha}) = \sum_{(u,i)\in \mathcal{O}} (r_{u,i} - q_i^T (\Psi_0 + \sum\limits_{(u,i) \in \mathcal{O}} \alpha_i \Psi_i^{r_{u,i}}) )^2 + \lambda |\mathbf{\alpha}|
\label{eqiam_cold}
\end{equation}
which is now continuous. Note that, in that case, the translation resulting from a rating over an item  corresponds to $\alpha_i \Psi_i^{r_{u,i}}$ rather than to $\Psi_i^{r_{u,i}}$. \\ \hfill~\linebreak
This objective loss can be optimized by using stochastic gradient-descent methods. Since it contains a $L_1$ term which is not derivable on all the points, we propose to use the same idea than proposed in \cite{carpenter2008lazy} which consists in first making a gradient step without considering the $L_1$ term, and then applying the $L_1$ penalty to the weight to the extent that it does not change its sign. In other words, a weight $\alpha_i$ is clipped when it crosses zero. 
%
%


\subsection{IAM and Classical Collaborative Filtering}

The IAM, which is particularly well-fitted for user cold-start recommendation, can also be used in the classical collaborative filtering problem, without constraining the set of items. In that case, the objective function can be written as:
\begin{equation}
\mathcal{L}^{warm}(\mathbf{q},\Psi) = \sum_{(u,i)\in \mathcal{O}} (r_{u,i} - q_i^T (\Psi_0 + \sum\limits_{(u,i) \in \mathcal{O}} \Psi_i^{r_{u,i}}) )^2
\label{eqiam_cf}
\end{equation}
which can be easily optimized through gradient descent. This model is a simple alternative to matrix factorization-based approaches, which is also evaluated in the experimental section. This model have some nice properties in comparison to transductive techniques, mainly it can easily update users' representations when faced with new incoming ratings, but this is not the topic of this article.

\section{Experiments}
\label{exp}

\begin{table}[t]
\vspace{-0.5cm}
\begin{subfigure}{.5\textwidth}
\begin{center}
\small{
\begin{tabular}{|c|c|c|c|} \hline 
DataSet & Users &  Items & Ratings  \\ \hline \hline
\textbf{ML1M} & 5,954 & 2,955 & 991,656 \\ \hline
%
\textbf{Flixter} & 35,657 & 10,247 &  7,499,706 \\\hline
%
\textbf{Jester} & 48,481 & 100 & 3,519,324  \\\hline
%
\textbf{Yahoo} & 15,397 & 1000 & 311,672 \\\hline
\end{tabular}
}
\end{center}
\caption{Description of the datasets}
\label{tab:datasets}
\end{subfigure}
\begin{subfigure}{.5\textwidth}
\begin{center}
\small{

\begin{tabular}{|c|c|c||c|} \hline 
     DataSet    & MF & IAM & ItemKNN \\\hline 
Jester  & 0.723       &  \textbf{0.737}   & 0.725 \\ 
ML1M  &  0.689    &  \textbf{0.727} & 0.675 \\ 
Yahoo   & 0.675   &  0.719         &  \textbf{0.726}\\ 
Flixter   & \textbf{0.766}	  &  0.758   &  NA  \\ 
\hline
\end{tabular}

}
\end{center}
\caption{Accuracy of the different models in the classical CF context (without cold-start). NA (Not Available) means that, due to the complexity of ItemKNN, results were not computed over the Flixter dataset.}
\label{tab:IAM_accuracy}
\end{subfigure}
\caption{Datasets description and performance of different models.}
\vspace{-0.5cm}
\end{table}

\begin{figure}[t]
\vspace{-0.5cm}
\centering
\begin{subfigure}{.5\textwidth}
  \centering
  \includegraphics[width=0.9\linewidth]{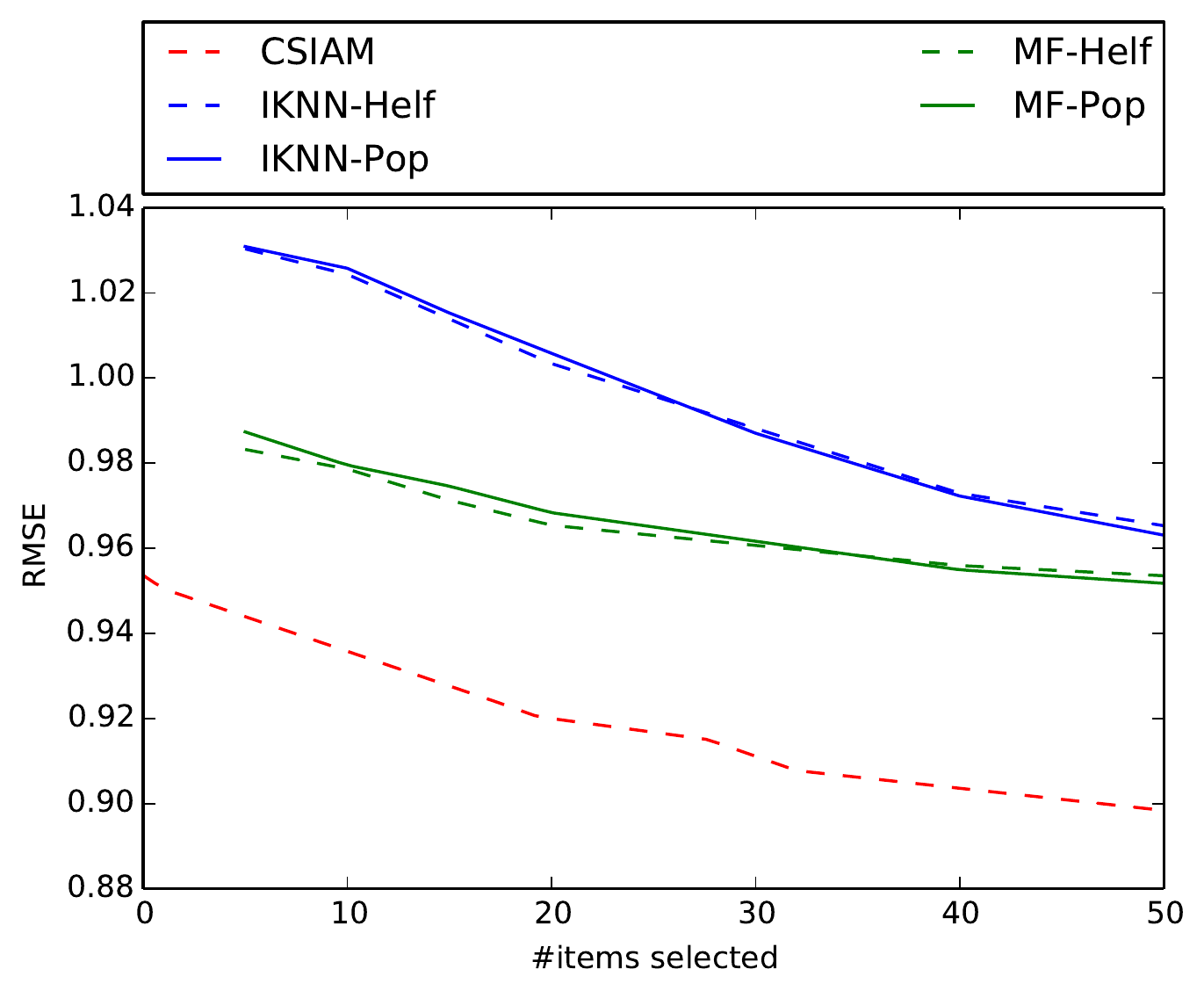}
  \caption{RMSE performance}
  \label{fig:yahoo_RMSE}
\end{subfigure}%
\begin{subfigure}{.5\textwidth}
  \centering
  \includegraphics[width=0.9\linewidth]{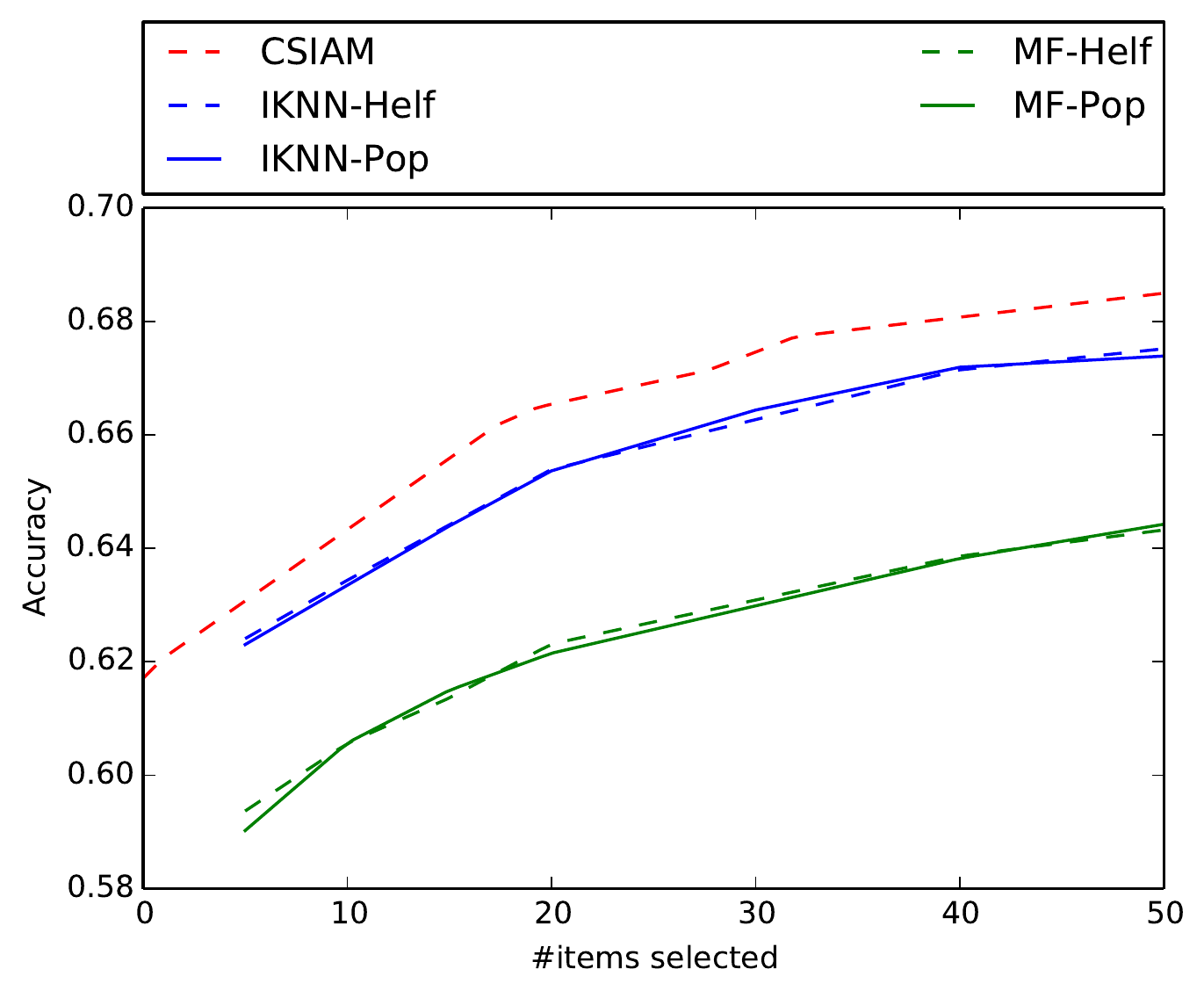}
  \caption{Accuracy performance}
  \label{fig:yahoo_Accu}
\end{subfigure}
\caption{Accuracy and RMSE evaluation on \textbf{Yahoo} dataset for all models, regarding the size of the interview (number of questions/items asked). }
\vspace{-0.5cm}
\label{fig:yahoo_perf}
\end{figure}

We evaluate our models on four benchmark datasets - Table \ref{tab:datasets} - of various size in terms of number of users, of items or regarding the sparsity of ratings. The datasets are classical datasets used in the literature (\cite{zhou2011functional,golbandi2010bootstrapping}). \textbf{ML1M} corresponds to the \textit{MovieLens 1 millon} dataset and \textbf{Yahoo} corresponds to the \textit{Yahoo! Music} benchmark. \textbf{Flixter} and \textbf{Jester} are classical datasets. As our main goal is mainly to evaluate the quality of our approach in the context of \textit{new users} arriving in the system, we define the following protocol in order to simulate a realistic interview process on incoming users, and to evaluate different models. We proceed as follow: (i) We randomly divide each dataset along users, to have a pool of \textit{training} users denoted $\mathcal{U}^{train}$, composed of 50\% of the users of the complete dataset, on which we learn our model. The remaining users are split in two sets (
representing each 25\% of initial users) for validation and testing. The interview process will be applied on each of these two subsets. (ii) The $\mathcal{U}^{test}$ and $\mathcal{U}^{valid}$  sets are then randomly split in two subsets of ratings to simulate the possible known answers : 50\% of the ratings of a set are used as the possible answers to the interview questions (\textit{Answer Set}). The 50\% of ratings left will be used for evaluating our models (\textit{Evaluation Set}).  Ratings have been binarized for each datasets, a rating of -1 (resp. 1) being considered a dislike, (resp. like).

%
The quality of the different models is evaluated by two different measures. The \textit{root mean squared error} (RMSE) measures the average ratings' prediction precision measured as the difference between predicted and actual ratings $(\hat{r}_{u,i} - r_{u,i})^2$.  
As we work with binary ratings, we also use the accuracy as a performance evaluation. In this context, it means that we focus on the overall prediction, i.e on the fact that the system has rightly predicted \textit{like} or \textit{dislike}, rather than on its precision regarding the "true" rating. The accuracy is calculated as the average "local" accuracy along users. These measures are computed over the set of missing ratings i.e the \textit{Evaluation Set}.


We explore the quality of our approach on both the classical CF context using the IAM Model (Equation \ref{eqiam_cf}) and on the cold-start problem using the CS-IAM model defined in Equation \ref{eqiam_cold}.  We compare our models with two baseline collaborative filtering methods: Matrix Factorization (MF) that we presented earlier, and the Item-KNN with Pearson correlation measure (\cite{koren2010factor}) which does not compute representations for users nor items but is a state-of-the-art CF method. Note that the inductive models (IAM and CS-IAM) are trained using only the set of training users $\mathcal{U}^{train}$. The ratings in the \textit{Answer Set} are only used as inputs during the testing phase, but not during training. Transductive models are trained using both the training users $\mathcal{U}^{train}$, but also the \textit{Answer set} of ratings defined over the testing users. 
It is a crucial difference as our model has significantly less information during training. 

\begin{figure}[t]
\vspace{-0.5cm}
\begin{subfigure}{0.4\textwidth}
  \centering
  \includegraphics[width=\linewidth]{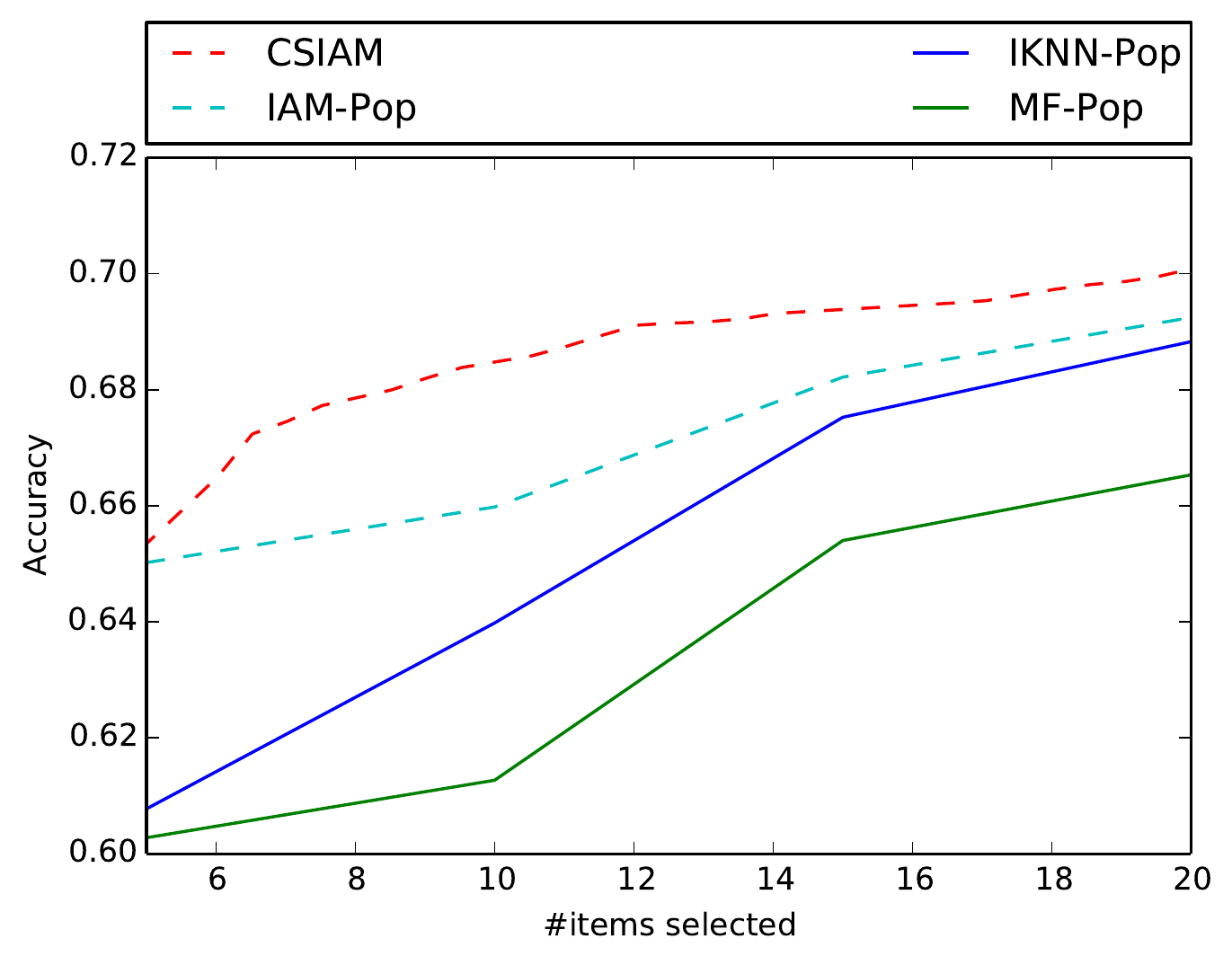}
  \caption{Accuracy performance on \textbf{Jester}}
  \label{fig:jester_pop}
\end{subfigure}
\begin{subfigure}{0.6\textwidth}
  \centering
	\includegraphics[width=0.8\linewidth]{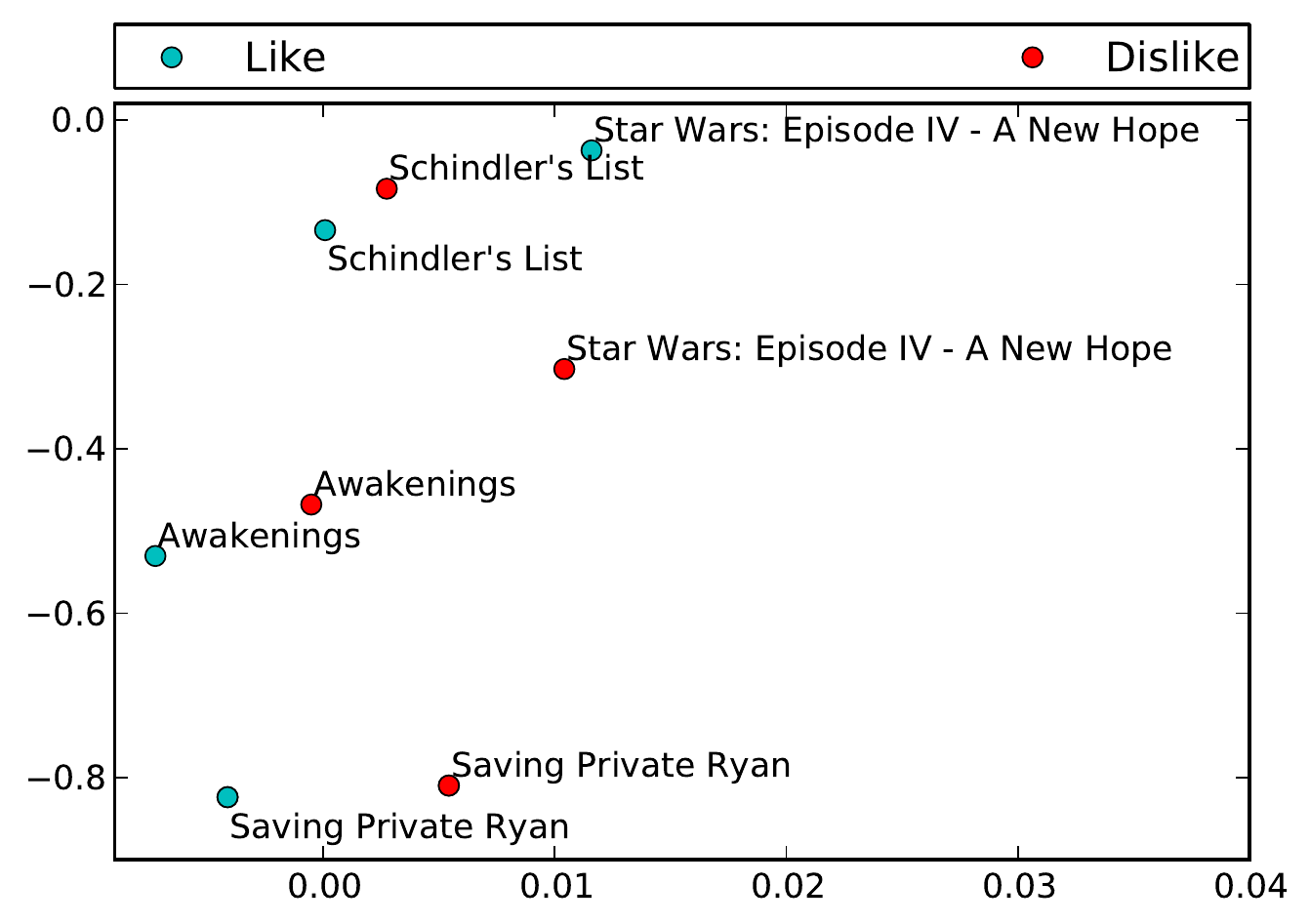}
  \caption{Visualization of some $\alpha_i \Psi_i$ after a PCA}
  \label{fig:2d}
\end{subfigure}
\caption{Performance on Jester, and visualization}
\vspace{-0.5cm}
\end{figure}

Each model has its own hyper-parameters to be tuned: the learning-rate of the gradient descent procedure, the size $N$ of the latent space, the different regularization coefficients... The evaluation is thus made as follows: models are evaluated for several hyper-parameters values using a grid-search procedure, the performance being averaged over 3 different randomly initialized runs. The models with the best average performance are presented in the next figures and tables. All models have been evaluated over the same datasets splits.

\subsection{Collaborative Filtering}
First, we evaluate the ability of our model to learn relevant representations in a classical CF context. In that case, the IAM model directly predicts ratings based on the ratings provided by a user. Results for the four different datasets are presented in Table \ref{tab:IAM_accuracy}. We can observe that, despite having much less information during the learning phase, IAM obtains competitive results, attesting the ability of the additive model to generalize to new users. More precisely, IAM is better than MF on three out of four datasets. For example, on the MovieLens-1M dataset, IAM obtains 72.7\% in terms of accuracy while MF's accuracy is only 68.9\%. Similar scores are observed for Jester and Yahoo. Although Item-KNN model gives slightly better results for two datasets, one should note that this method do not rely on nor provide any representations for users or items and belongs to a different family of approach. Moreover, ItemKNN - which is based on a KNN-based method - has a high complexity, and is 
thus very slow to use, and unable to deal with large scale datasets like Flixter on which many days are needed in order to compute performance.  Beyond its nice performance IAM is able to predict over a new user in a very short-time, on the contrary to MF and ItemKNN.




\subsection{Cold-start Setting}

We now study the ability of our approach to predict ratings in a realistic cold-start situation. As MF and ItemKNN do not provide a way to select a set of items for the interview, we use two benchmark selection methods used in the literature (\cite{rashid2002getting}). The \textbf{POP} method select the most popular items - i.e the items with the highest number of ratings in the training set - and the \textbf{HELF} (\textit{Harmonic mean of Entropy and Logarithm of rating Frequency}) method which select items based on both their popularity but also using an entropy criterion, which focus on the \textit{informativeness} of items (e.g a controversial movie can be more informative than a movie liked by everyone) (\cite{rashid2008learning}). Our model is learned solely on the $\mathcal{U}^{train}$  set. Baselines are computed on a dataset composed of the original $\mathcal{U}^{train}$  ratings with the additional ratings of the \textit{AnswerSet} of $\mathcal{U}^{test}$  that lie into the set of 
items selected by the POP or the HELF approach. Transductive approaches use more information during training that our inductive model. 

The number of items selected by the CS-IAM model directly depends on the value of the $L_1$ regularization coefficient and several values have been evaluated.  In CS-IAM, the number of selected items correspond to the number of non-null $\alpha_i$ parameters. The number of items selected by POP and HELF is manually chosen.

Figure \ref{fig:yahoo_perf} shows accuracy and RMSE results for all models on the Yahoo dataset as a function of the interview size. It first illustrates that ItemKNN approach does not provide good results for RMSE-evaluation, as it is not a \textit{regression}-based method, but is better than MF in terms of accuracy. It also shows that HELF criterion does not seem to be specifically better on this dataset than the POP criterion. For both evaluations, CS-IAM gives better results, for all sizes of interview. It can also be noted that CS-IAM also gives good results when no item is selected due to the $\Psi_0$ parameters that correspond to the learned default representation. The model with $0$ items also expresses the base performance obtained on users unable to provide ratings during the interview. 

Detailed accuracy results for the four datasets are summarized in Table \ref{tab:BIAM_accuracy}, for different reasonable sizes of interview. Similar observations can be made on the results, 
where CS-IAM managed to have the best or competitive accuracy for all datasets and all number of questions allowed, while using less information in train.



At last, when comparing the performance of CS-IAM with a version of IAM where items have been selected by the POP criterion -IAM-Pop, Figure \ref{fig:jester_pop} - one can see that the CS-IAM outperforms the other approaches. It interestingly shows that (i) IAM managed to give better results than MF with the same information selection strategy (POP) (ii) CS-IAM with all its parameters learned, managed to select more useful items for the interview process, illustrating that the performance of this model is due to both, its expressive power, but also on its ability to simultaneously learn representations, and select relevant items.


\begin{table}[t]
\vspace{-0.5cm}
\begin{center}
\small{
\begin{tabular}{|c|c|c|c|c|c|c|}
	\cline{1-7}
	DataSet & NbItems & MF POP & MF HELF & IKNN POP & IKNN HELF & CS-IAM \\\hline\hline
    
	\multicolumn{1}{ |c  }{\multirow{4}{*}{Jester} } &
	
	\multicolumn{1}{ |c| }{5} & 0.603 & 0.589 & 0.608 & 0.634 & \textbf{ 0.667 } \\ 
	\multicolumn{1}{ |c  }{}    &
	\multicolumn{1}{ |c| }{10} & 0.613 & 0.609 & 0.640 & 0.608 & \textbf{ 0.686 }  \\ 
	\multicolumn{1}{ |c  }{}                        &
	\multicolumn{1}{ |c| }{20} & 0.665 & 0.641 & 0.688 & 0.676 &  \textbf{0.701 } \\ 
\hline\hline
    

	\multicolumn{1}{ |c  }{\multirow{4}{*}{MovieLens 1M} } &
	
	\multicolumn{1}{ |c| }{5} & 0.629 & 0.617 & 0.649 & 0.647 & \textbf{ 0.690  } \\ 
	\multicolumn{1}{ |c  }{}  &
	\multicolumn{1}{ |c| }{10} & 0.634 & 0.620 & 0.651 & 0.653  &  \textbf{ 0.695 }  \\ 
	\multicolumn{1}{ |c  }{}                        &
	\multicolumn{1}{ |c| }{20} & 0.648 & 0.621 & 0.663 & 0.638 &  \textbf{ 0.696  }  \\ 
\hline\hline
    

	\multicolumn{1}{ |c  }{\multirow{4}{*}{Yahoo} } &
	
	\multicolumn{1}{ |c| }{5} & 0.590 & 0.594 & 0.623 & 0.624 &\textbf{ 0.638 }  \\ 
	\multicolumn{1}{ |c  }{}  &
	\multicolumn{1}{ |c| }{10} & 0.601 & 0.610 & 0.633 & 0.634 & \textbf{ 0.647 } \\ 
	\multicolumn{1}{ |c  }{}                        &
	\multicolumn{1}{ |c| }{20} & 0.621 & 0.623 & 0.654 & 0.654 &  \textbf{0.665}    \\ 
\hline\hline
      
\multicolumn{1}{ |c  }{\multirow{4}{*}{Flixter} } &
	
	\multicolumn{1}{ |c| }{5} & 0.719 & 0.722 & NA & NA & \textbf{0.723}   \\ 
	\multicolumn{1}{ |c  }{}  &
	\multicolumn{1}{ |c| }{10} & 0.720 & 0.726 & NA & NA & \textbf{0.727}   \\ 
	\multicolumn{1}{ |c  }{}                        &
	\multicolumn{1}{ |c| }{20} & 0.727 & \textbf{0.739} & NA  & NA & 0.735   \\ 
    
\hline
\end{tabular}
}
\end{center}
\caption{Accuracy performance of models on four datasets regarding the number of questions asked. NA (Not Available) means that, due to the complexity of ItemKNN, results were not computed over the Flixter dataset. Bold results corresponds to best accuracy.\vspace{-0.6cm}}
\label{tab:BIAM_accuracy}
\end{table} 

We have shown that our approach gives significantly good quantitative results. We now focus our interest on a \textit{qualitative} analysis of the results performed over the MovieLens dataset. First, we compare the items selected by the three selection methods (CS-IAM, POP and HELF). These items are presented in Table \ref{tab:ml1M_interview}. First, when using the POP criterion, one can see that many redundant movies are selected - i.e the three last episodes of Star Wars on which the ratings are highly correlated: a user likes or  dislikes Star Wars, not only some episodes. The same effect seems to appear also with CS-IAM which selects \textit{Back to the future I} and \textit{Back to the future III}. But, in fact, the situation is different since the ratings on these two movies have less correlations. Half of the users that like \textit{Back to the future I} dislike \textit{Back to the future III}. 

Figure \ref{fig:2d} shows the translations $\alpha_i\Psi_i$ after having performed a PCA in order to obtain 2D representations. What we can see is that depending on the movie, the fact of having a positive rating or a negative rating does not have the same consequences in term of representation:  For example, liking or disliking \textit{Saving Private Ryan} is different than liking or disliking \textit{Star Wars}; the translation concerning these two movies are almost perpendicular and thus result in a very different modification of the representation of the user. \textit{Schindler's List} has less consequences concerning the user representation i.e the norm of $\alpha_i \Psi_i^r$ is lower than the others. 


\subsection{Mixing Cold-start and Warm Recommendation}

Our model can also allow one to smoothly move from a cold-start to a \textit{warm} context : after having answered the interview, the user will start interacting with the system, providing new ratings, which will be easily integrated with our inductive translation model to update his representation and thus, the resulting recommendations. To do so, we simply change the learning strategy: (i) The model is learned in the warm setting described in Equation (8), i.e we learn each item's representation $q_i$ and the translations on representations (the $\Psi_i^r$ parameters). (ii) We select the most relevant items for the interview process by learning the $\alpha$'s weights using a $L_1$ regularization as explained in Equation (7). In this phase, we only learn the $\alpha$-values which will allow us to choose which items to use during the interview, following Equation (6). After the interview, each new incoming rating modifies the user representation as explained in Equation (4), resulting in a system that is naturally able to take into account new information.  Note that, in this setting, the use of an hyperbolic tangent function on the representation, which will limit its norm, improves the quality of the system.

\begin{figure}[t]
  \begin{minipage}[b]{0.4\textwidth}
	\centering
       \includegraphics[width=\linewidth]{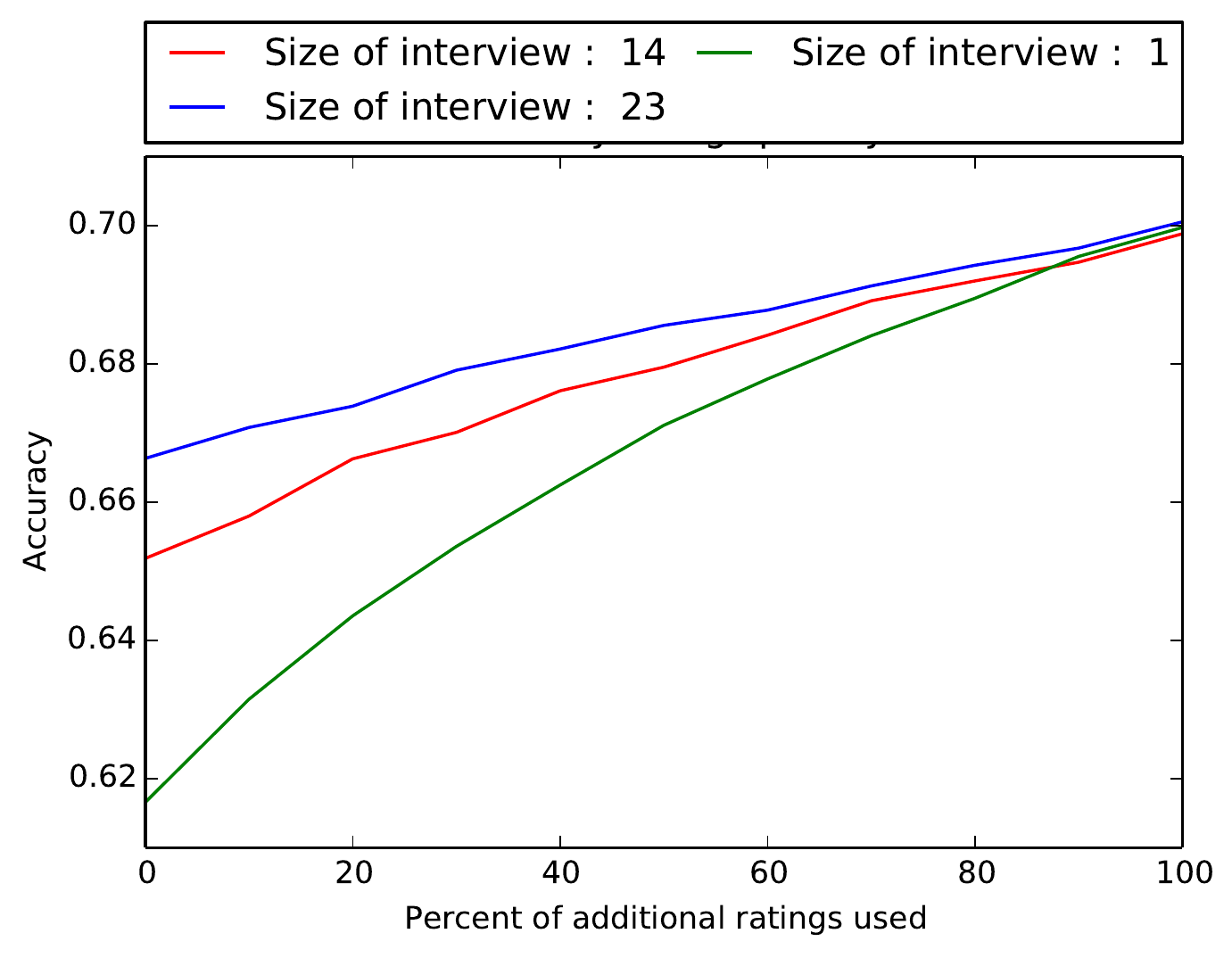}
      \captionof{figure}{Accuracy regarding percentage of ratings added after the interview}
  \label{fig:sparsity_yahoo}
  \end{minipage}
  \begin{minipage}[b]{0.60\textwidth}
    \centering
    
\small{
\begin{tabular}{|p{3.4cm}|p{3.4cm}|} \hline 
\textbf{CS-IAM} & \textbf{Popularity}\\ \hline
American Beauty, Being John Malkovich, Lion King, Ghost, Superman, Back to the Future, Fargo, Armageddon, Get Shorty, Splash, 20 000 Leagues Under the Sea, Back to the Future Part III, Outbreak & 
American Beauty, Star Wars: Episode I,   Star Wars: Episode V, Star Wars: Episode IV, Star Wars: Episode VI, Jurassic Park,  Terminator 2,  Matrix,  Back to the Future,  Saving Private Ryan, Silence of the Lambs, Men in Black, Raiders of the Lost Ark, Fargo  \\ \hline
\end{tabular}
}
\captionof{table}{MovieLens 1M - Selected items for the interview process by the three selection methods }
\label{tab:ml1M_interview}

    \end{minipage}
  \end{figure}

  This model has been evaluated on the \textbf{Yahoo} dataset with the following experimental protocol: First the model is evaluated in its cold-start setting using the item with non-null $\alpha$'s values. Then, we evaluate the performance of this model when adding different amount of ''new'' ratings sampled uniformly from the set of items. The results are illustrated in Figure \ref{fig:sparsity_yahoo} which shows that the performance of this strategy increases as new ratings are added and almost reaches the one obtain for the classical warm setting (see Table \ref{tab:IAM_accuracy}). Curves for different sizes of initial interviews are shown. We think that this extension of our approach which makes the link between the cold-start and the warm settings is an original and promising feature.

~\\

\vspace{-0.6cm}
\section{Related Work}
\label{rw}

The recommendation problem has been studied under various assumptions. We focus on \textbf{Collaborative Filtering} (CF) methods, which only use the past ratings observed on the users and items, but other families of approaches exists, as \textit{Content-Based} methods, which use informative features on users and items (\cite{pazzani2007content}), and hybrid methods that mix ratings and informative features (\cite{basilico2004unifying}).
\\CF techniques can be distinguished into two categories. \textit{Memory-based} methods, such as Neighbor-based CF \cite{resnick1994grouplens},  calculate \textit{weights} between pairs of items (\cite{sarwar2001item}) or users (\cite{herlocker1999algorithmic}), based on similarities or correlations between them. \textit{Model-based} methods, such as \textit{Latent Factor Models}, have rather a \textit{representation learning} approach, where representations vectors for each user and item are inferred from the matrix of ratings with matrix factorization techniques (\cite{koren2009matrix}). 
 Collaborative filtering models have a major limitation when there is no history for a user or an item. A classical approach in this case is to use an interview process with a few questions asked to the new user as it is done in this paper. Several papers have proposed different methods to choose which questions to select. \textit{Static approaches} (see \cite{rashid2002getting} for a comparative study), construct a static seed set of questions (fixed for all users) following a selection criterion like measures of popularity, entropy or coverage while \cite{golbandi2010bootstrapping} also proposed a greedy algorithm that aims to minimize the prediction error performed with the seed set. \textit{Adaptive approaches} have also been proposed, where the interview process considers the user's answers to choose the next question. For example, \cite{rashid2008learning} fits a decision tree to find a set of clusters of users, while \cite{golbandi2011adaptive} uses a ternary tree where each node is an item and branch corresponds to eventual answers (\textit{like,dislike,unknown}). \cite{zhou2011functional} presents \textit{functional matrix factorization}, a decision tree based method which also associate a \textit{latent profile} to each nodes of the tree. 
The closest model to our approach is \cite{sun2013learning}, who learn a ternary tree allowing multiple questions at each node, each node containing a (learned) regressor and translations functions on selected items. Our model can be seen as one node of their tree. However, their approach does not seem to allow a bridge between cold start and warm context as ours does. 
It is also interesting to note that while usually more efficient, one drawback of such adaptive approaches is that users usually dislike having to rate item one by one and prefer rating several items in one shot (\cite{golbandi2011adaptive,rashid2002getting}).
 

\section{Conclusion and Perspectives}
\label{conclusion}
We have proposed a new representation-based model for collaborative filtering. This inductive model (IAM) directly computes the representation of a user by cumulative translations in the latent space, each translation depending on a rating value on a particular item. We have also proposed a generic formulation of the user cold-start problem as a representation learning problem and shown that the IAM method can be instantiated in this framework allowing one to learn both which items to use in order to build a preliminary interview for incoming users, but also how to use these ratings for recommendation. The results obtained over four datasets show the ability of our approach to outperform baseline methods. 
Different research directions are opened by this work: (i) first, the model can certainly be extended to deal with both incoming users, but also new 
items. In that last case, the interview process would consist in asking reviews for any new item to a particular subset of relevant users. (ii) While we have studied the problem of building a static interview - i.e the opinions on a fixed set of items is asked to any new user - we are currently investigating how to produce personalized interviews by using sequential learning models i.e reinforcement learning techniques.

\section*{Acknowledgements}\vspace{-0.4cm}
This article has been supported within the Labex SMART supported by French state funds managed by the ANR within the Investissements d'Avenir programme under reference  ANR-11-LABX-65. Part of this work has benefited from a grant from program DGA-RAPID, project LuxidX.
\vspace{-0.4cm}
\bibliography{biblio}

\bibliographystyle{iclr2015}
\end{document}